# Plasma panel-based radiation detectors

**Peter Friedman** (SID Member)
**Robert Ball**
**James Beene**
**Yan Benhammou**
**Meny Ben-Moshe**
**Hassan Bentefour**
**J. W. Chapman**
**Erez Etzion**
**Claudio Ferretti**
**Daniel Levin**
**Yiftah Silver**
**Robert Varner**
**Curtis Weaverdyck**
**Bing Zhou**

*Abstract* — The plasma panel sensor (PPS) is a gaseous micropattern radiation detector under current development. It has many operational and fabrication principles common to plasma display panels. It comprises a dense matrix of small, gas plasma discharge cells within a hermetically sealed panel. As in plasma display panels, it uses nonreactive, intrinsically radiation-hard materials such as glass substrates, refractory metal electrodes, and mostly inert gas mixtures. We are developing these devices primarily as thin, low-mass detectors with gas gaps from a few hundred microns to a few millimeters. The PPS is a high gain, inherently digital device with the potential for fast response times, fine position resolution (<50-µm RMS) and low cost. In this paper, we report on prototype PPS experimental results in detecting betas, protons, and cosmic muons, and we extrapolate on the PPS potential for applications including the detection of alphas, heavy ions at low-to-medium energy, thermal neutrons, and X-rays.

DOI # 10.1002/jsid.151

## 1 Introduction

The plasma panel sensor (PPS) is a new radiation detector technology being developed for a number of scientific and commercial applications.[1–5] The PPS (Fig. 1), which is based on the plasma display panel (PDP), is designed to exploit low-cost fabrication methods employed for PDPs and liquid crystal displays in HDTVs. PDPs are composed of millions of cells per square meter, each of which when provided with a signal pulse can initiate and sustain a plasma discharge. Rather than the plasma discharge being initiated *externally* by a signal from a driver chip, the PPS Geiger-mode discharge is initiated *internally* by ion pairs created within the device by an ionizing photon or particle interacting with the detector gas (or wall). The bias voltage across the cell is set to exceed the breakdown voltage, as described by Paschen's Law. The ionizing event creates an electron avalanche that ultimately results in a gaseous discharge whose amplitude is limited by the cell capacitance. The discharge is terminated by the presence of a localized quench resistance. An important operating principal of the quench resistance is that, combined with the cell capacitance, it yields an RC time constant or cell recovery time long enough that the free charges in the gas volume are neutralized and the metastable states deactivated. In summary, in a PPS, instead of applying voltage to produce light emission via a plasma discharge as in a PDP, we detect the discharge from an ionization event generated by radiation interacting with the cell media.

## 2 Experimental: panel structure and readout electronics

The PPS is conceived as a dense array of micro-Geiger cells, each having a discharge gap on the order of 100–500 µm. The configuration of a cell can have the discharge through the volume of the cell, a columnar-discharge structure (Fig. 1) or it can confine the discharge near a surface, a surface-discharge structure. In both cases, the discharge is initiated by electron–ion pairs produced in the gas volume of the cell. In the latter case, the ions must drift to the discharge surface. We anticipate that the size of the gas volume and the type of the discharge structure can be varied for different applications. For example, low-energy betas, protons, and radioactive ions can be highly ionizing, depending upon their energy, whereas relativistic muons are *minimum* ionizing particles (MIPS) and so require a much larger gas path (e.g., perhaps a few millimeters) to generate an equivalent number of ion pairs. It is the former configuration however that has been prototyped and predominately tested to date.

The active area in the columnar-discharge PPS structure consists primarily of the gas volume between the electrodes, which is enclosed by the front and back substrates. The test panel shown in Fig. 2 is a columnar-discharge, two-electrode, DC-mode (direct current), glass PPS in a removable aluminum frame, fitted with a sealed, high-vacuum, shut-off valve to allow multiple fills at different pressures and of different gas mixtures. This structure has proven exceedingly useful as we can hold a given gas mixture for at least 2 to 3 months without observing any change in performance. By using the same panel in this manner, we can study the effect of changes in gas mixture completely isolated from any uncertainty associated with panel-to-panel variations in discharge and gas gap, electrode linewidth and surface condition, dielectric wall structure, substrate thickness, pixel surface defects, and others. The panel in Fig. 2 has an active area of 8.1 × 32.5 cm, with a pixel/electrode pitch of 2.5 mm, and has been fabricated with either transparent $SnO_2$ or Ni column HV (high voltage) electrodes (i.e., cathodes), and Ni back row sense anodes. We have also fabricated similarly constructed panels with a pixel pitch of 1.0 mm. All panels undergo a systematic evacuation, bakeout, and gas-fill procedure before being operated as detectors. They have produced the gas discharge pulses and the data reported in this paper.

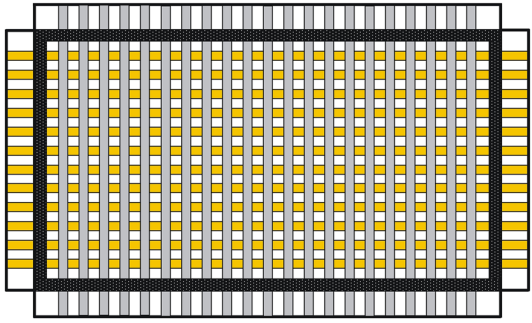

**FIGURE 1** — Drawing of two-electrode, columnar-discharge PPS structure. Orthogonal Ni (or $SnO_2$) electrodes are separated by a few hundred micron gas layer (see text and Fig. 2). The dark band around the perimeter is a hermetic glass seal.

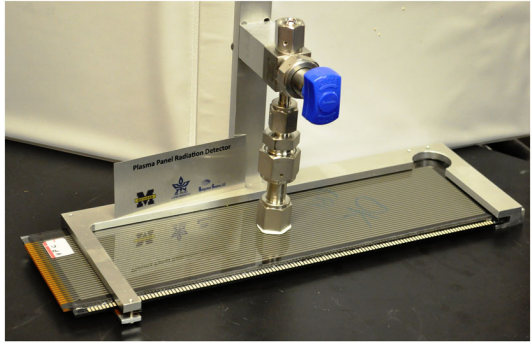

**FIGURE 2** — Modified DC-PDP columnar-discharge PPS test panel with "refillable" gas valve.

To explore the behavior of PPS devices under various kinds of radiation, we have constructed two test benches: one at the University of Michigan (U-M) and the other at Tel Aviv University. Each test bench includes a gas delivery system, several radiation sources, a triggering system, and a data acquisition (DAQ) system. We also have access to a ProCure medical proton beam accelerator near Chicago through an informal collaboration with Belgium proton beam therapy manufacturer Ion Beam Applications S.A. (IBA). We used their model C235 accelerator to test our devices with a 226-MeV collimated proton beam using aperture diameters of 1 and 10 mm. The triggering system for our lab-based experiments is carried out with a scintillator hodoscope (Fig. 3) or relies on self-triggering. The proton test beam data were acquired with a PPS self-trigger. The DAQ is for the characterization of the signal induced in the panel during discharge. To accomplish this, we are using two sets of 5-GHz digitizer boards (i.e., digital sampling oscilloscope) based on the DRS4 chip developed at the Paul Scherrer Institut (http://drs.web.psi.ch/). For the discharge rate measurements, we are using a set of discriminators and counters (Fig. 3). With the two digitizers (four channels each), we are able to read a 4 × 4 array of pixels simultaneously to achieve a 2D position measurement of radiation traversing the panel. More recently, we have transitioned to a DAQ adapted from the ATLAS Muon Spectrometer monitored drift tube readout electronics developed (in part by U-M) for the Large Hadron Collider. This effort has required the development or implementation of several subcomponents: (1) design and production of front-end signal pickoff cards; several iterations of these cards have been produced and currently provide good signal processing from the panels; (2) configuration of a VME-based (Versa Module Eurocard) DAQ system using specialized components obtained from CERN (European Organization for Nuclear Research); (3) DAQ software; and (4) development of data analysis software. The first generation of the new DAQ readout electronics has been completed with the capability to acquire PPS data for 24 multiple readout channels with nanosecond resolution. These electronics are now being modified and reconfigured to provide a much more compact and portable system.

To determine the panel response to radiation, we used GEANT4[6] to simulate the energy loss and scattering occurring in the glass substrate prior to low-energy beta particles entering the panel gas discharge region. The beta energy spectrum is based on the known energy originally emitted by both our $^{90}$Sr and $^{106}$Ru sources. Most of our efforts have focused on the response of columnar-discharge PPS devices (e.g., modified DC-PDPs), which produce signals when exposed to radioactive sources or when being traversed by a cosmic muon. The experiments described and the results reported here have been focused on the detection of charged particles by *direct* interaction and ionization of the gas.

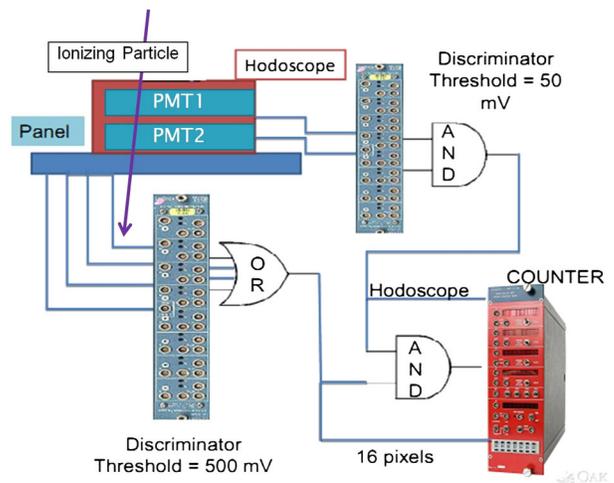

**FIGURE 3** — Hodoscope measurement setup for either double or triple coincidence discharges from $^{106}$Ru beta particles or from cosmic muons. Output signals are read out from the anodes from a 50Ω termination. Signals are typically attenuated.

## 3 Results and discussion

We have investigated the PPS device response to a number of ionizing particle sources under different experimental conditions with various discharge gases. The gas pressures have ranged from about 200 to 700 Torr. The observed signals from all of the devices tested have had large amplitudes of at least several volts, so there has been no need for amplification electronics. In fact, the discharge signals have typically been too large for our readout electronics, thus requiring significant attenuation. For each gas tested, the shape of the induced signals is uniform. The leading edge rise time for our current generation of panels is typically 1–2 ns (Fig. 4). The discharge spreading to neighboring pixels is gas dependent but has been measured, for example, to be ~2% for an Ar/$CO_2$ mixture to a single nearest-neighbor pixel.[4] A typical example is shown in Fig. 6, which is a single pixel response to a $^{106}$Ru beta source, with no response seen on the neighboring electrodes (i.e., the adjacent electrodes/channels are shown in different colors).

The PPS discharge gases tested include the following: Ar + $CO_2$, Ar + $CF_4$, $CF_4$, $SF_6$, and Xe. Not unexpectedly, the device performance has been shown to be very much gas dependent, with the breakdown voltages varying by more than 1000 V for different gas mixtures in the same panel. The discharge spreading to neighboring cells is also gas dependent, yet we have shown that gas discharges can be confined to a single cell with several gas mixtures showing minimal, if any, gas discharge spreading to adjacent cells. We consider it significant that in an "open" cell structure (e.g., Figs 1 and 2), we have demonstrated minimal discharge spreading, especially given that our devices operate in the Geiger mode, producing large amplitude, high gain discharges. The fact that this has been done *without* an internal barrier structure around each cell is particularly encouraging. But equally important is that we have not added a hydrocarbon quenching gas component that would certainly degrade in a plasma discharge environment. The elimination of hydrocarbon quenching gases is considered critical to realizing a stable, hermetically sealed PPS device without the cost, bulk, and complication of having to constantly exchange the gas as required in a number of other position-sensitive gaseous detectors.

A "typical" gas discharge pulse is shown in Fig. 4 from a panel similar to that in Fig. 2. The PPS in Fig. 4 was filled with 1% $CO_2$ in 99% Ar at 600 Torr and operated at 840 V. The experiment employed a beta source of $^{106}$Ru. The 20–80% rise time was 1.2 ns (<2 ns for 10–90%), with a 1.9-ns pulse duration (full width at half-maximum). Depending on the specific gas and panel structure, the signal amplitudes can range from a few volts to tens of volts. These large amplitudes reflect the fact that the effective discharge capacitance for these test panels includes contributions from neighboring electrodes. This effect is discussed in more detail at the end of this section in conjunction with our SPICE[7] modeling program.

We have run the majority of our PPS tests using primarily four different particle sources of radiation: ~0.5 to 1 MeV betas from $^{90}$Sr (i.e., maximum energy of 2.3 MeV), higher energy betas from $^{106}$Ru (maximum energy of 3.5 MeV), relativistic particles/energies from cosmic muons (~4 GeV at sea level), and 226 MeV protons from an IBA-C235 accelerator. In all cases, the actual signal pulses look remarkably similar (e.g., see Fig. 4) for a given panel geometry, gas mixture, cathode voltage, and quench and signal resistors. In other words, the signal amplitude, rise time, and duration do not depend on the event causing the initial gas ionization. There is nothing surprising about this observation, as the cells are being driven in the Geiger or gas breakdown mode.

We present in Fig. 5 a plot of the cell count rate in hertz (10 min/point) versus high voltage for hits detected by four single cells using a 2.5-mm radius collimated $^{106}$Ru source in a PPS similar to that described earlier in Fig. 4. The panel in Fig. 5 was fabricated with Ni electrodes for both the HV (cathodes) and sense (anodes) electrodes and employed a gas mixture of 99% Ar/1% $CO_2$ at 600 Torr. As can be seen, the total number of background counts (i.e., without the source present) shown in "green" is near "zero" for most of the range, rising to a few hertz above 950 V. This shows that over a large range of signal producing voltages, the background rate is minimal. This behavior is similar to the results reported previously using a panel with transparent $SnO_2$ cathodes and filled with $CF_4$ at 500 Torr.[5] Although low background count rates in the absence of an efficiency measurement can be misleading, we consider the measured low rates to be a promising indication of good performance.

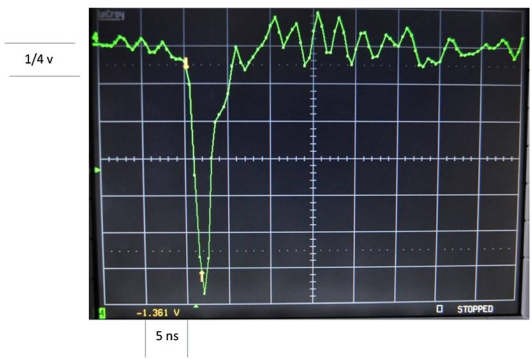

**FIGURE 4** — Typical pulse rise time and duration for a two-electrode, columnar-discharge PPS. Amplitudes depend on panel gap and gas type.

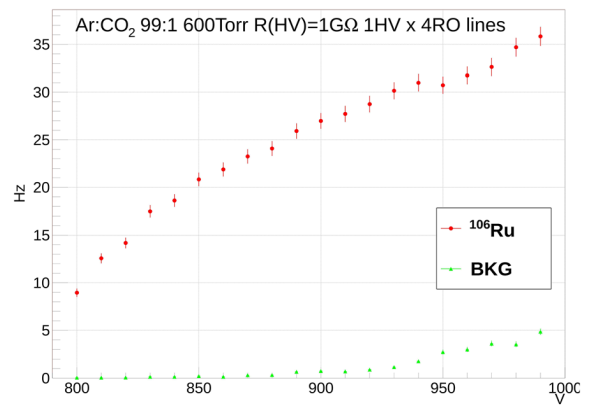

**FIGURE 5** — Example of hit rate versus high voltage of combined four pixels exposed to a $^{106}$Ru beta source. Background is measured in the absence of the source.

We show in Fig. 6 a gas discharge pulse (i.e., "blue" readout line #9) from an all-Ni, modified PDP filled with 600 Torr of 100% Xe. The source was $^{106}$Ru (beta source). The adjacent anode wires (i.e., channels 6, 7, and 8) appear as the black, red, and green lines, and show no indication of any discharge spreading.

The cosmic ray muon pulse generation time distribution is shown in Fig. 7 for the PPS in Fig. 2 filled with $SF_6$ at 200 Torr and operating at 1530 V. These data were for 197 muons collected during an overnight run using the scintillator hodoscope trigger. The pulse arrival times are relative to the hodoscope trigger with most circuit and cable delays removed. The data are nicely fit by a Gaussian with a width ($\sigma$) of less than 5 ns (i.e., arrival timing jitter) and a mean value of ~10 ns representing the average pulse generation formative time lag. Both $CF_4$ and $SF_6$ gases show similar response time signals of a few nanoseconds.

Figure 8 shows the translation of a $^{106}$Ru beta-source "collimated" through a 1.25-mm-wide graphite slit (20 mm thick) in 0.5-mm increments across the PPS sense electrodes (i.e., back plate row electrodes in Figs 1 and 2), with 1% $CO_2$ in 99% Ar at 600 Torr and 890 V. The plot shows the Gaussian means versus the source position. The mean (RMS) position resolution is ~0.7 mm, in a PPS panel with a 2.5-mm electrode pitch, which is encouraging given the incident dispersion of the betas going through the 2.25-mm-thick glass front substrate (Fig. 9). We obtain a slope of $0.39 \pm 0.01$ mm$^{-1}$ where the error is estimated from fitting the plot over three ranges. This slope is consistent with the electrode pitch.

A GEANT4 simulation was run of the beta scattering from a pencil beam of beta particles emanating out of the $^{106}$Ru source and traveling through the 20-mm-long air gap of the 1.25-mm-wide graphite collimator and then through the 2.25-mm-thick glass substrates of the PPS. The simulation also included the scattering contribution of the beta particles through the 0.44-mm path length of Ar discharge gas at 760 Torr. A total of 1,000,000 tracks were run for the GEANT4 simulation, with a computer-generated representation of 100 random tracks shown in Fig. 9. As can be seen in Fig. 9, most of the scattering and absorption of betas occur in the PPS front glass substrate with very few betas exiting the back glass substrate. This is why our coincidence experiments could not be easily performed using the lower-energy $^{90}$Sr beta source. Even with higher energy betas from the $^{106}$Ru source, significant time is required to accumulate a statistically reproducible signal. This is one reason why cosmic-ray muons and accelerated protons are so useful for this type of experiment, as the much higher energy of these particles is more than sufficient to penetrate the scintillator and glass layers, although for cosmic muons, the time required is very long due to their low intensity.

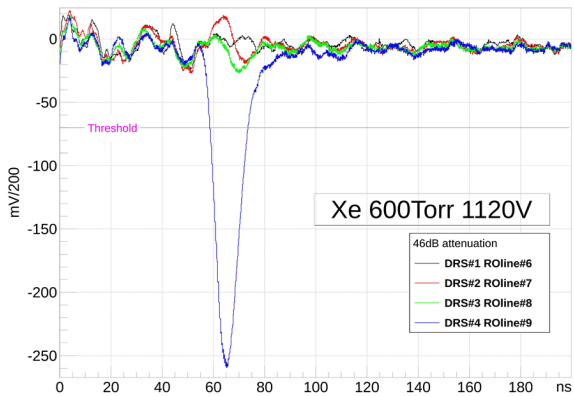

**FIGURE 6** — Discharge spreading experiment. The hit channel (blue) and nearest neighbors are shown.

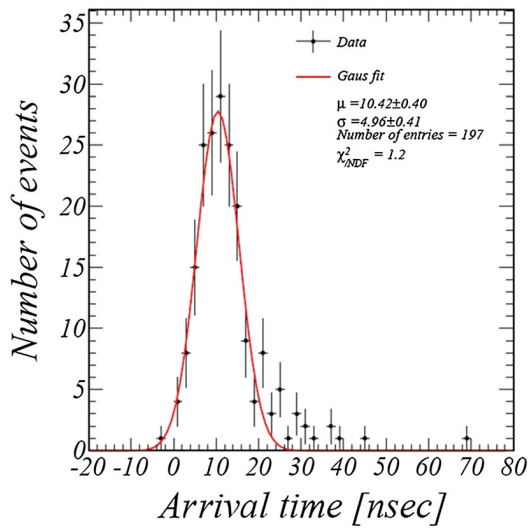

**FIGURE 7** — Temporal response (i.e., arrival time distribution) to cosmic ray muons using pure $SF_6$ gas at 200 Torr. The offset reflects residual cable delays. The timing jitter ($\sigma$) is 5 ns.

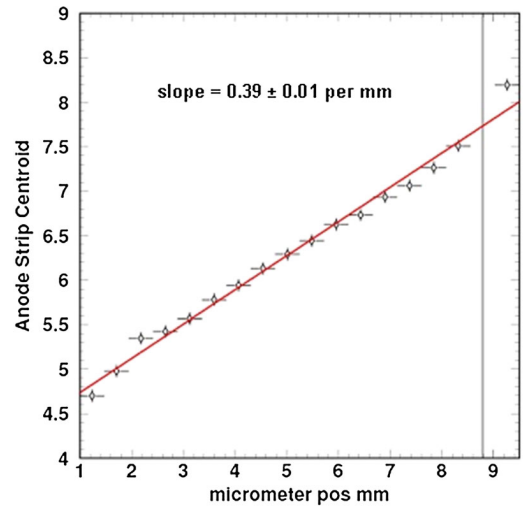

**FIGURE 8** — Position resolution measurements.

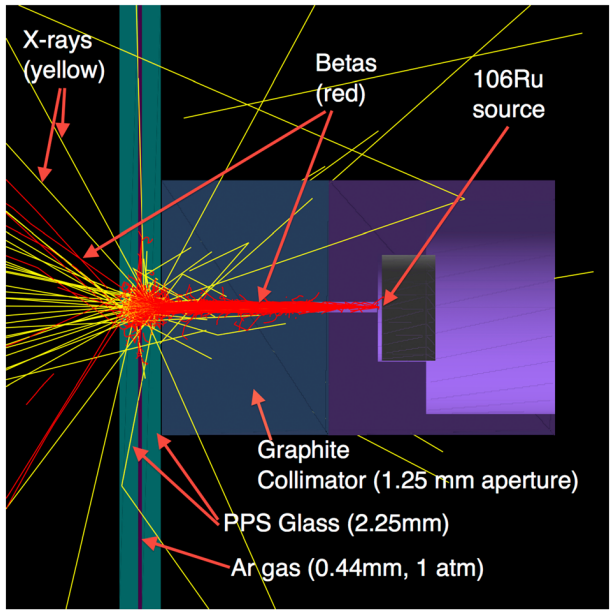

**FIGURE 9** — GEANT4 beta scattering simulation.

From the simulation in Fig. 9, we can see that a significant number of X-rays are also generated via inelastic scattering from the incident particle beam. These secondary X-ray emissions are also capable of generating a PPS discharge signal; however, their contribution should be small because of both their reduced number (i.e., in comparison with the betas) and their much lower probability of interaction with the Ar gas. Finally, we can see from Fig. 9 that the actual width of a pencil beam of beta particles by the time it reaches the discharge gas is at least several times its initial width. In other words, the "collimated" beta beam inside the PPS is actually distributed over one or two adjacent sense electrodes on each side of the targeted electrode under the graphite slit. Given this incident particle distribution, the fact that we are able to resolve the beam centroid to within ~0.7 mm in a PPS with an electrode pitch of 2.5 mm and a substrate thickness of 2.25 mm bodes very well for the potential position resolution of these devices. In this regard, we are currently fabricating PPS devices with a substrate thickness of 0.38 mm and planning to fabricate such devices with an electrode pitch of ~0.15 mm in the near future. We expect that such PPS devices should have a position resolution of better than 50 µm.

We performed our first particle beam experiments using an IBA-C235 proton beam accelerator used for the treatment of cancer. In Fig. 10 (right), we show the position scan using a 1-mm diameter, 226 MeV intense (>MHz/cm$^2$) proton beam for 16 sequential runs in which the panel in Fig. 2 was shifted in each run by ~1 mm increments relative to the fixed position proton beam. Each bin is a single data channel for a sense-electrode line. Figure 10 (left) shows the reconstructed position centroid of the "hit" map from Fig. 10 (right) versus the PPS relative displacement with respect to the initial position. The position centroid for each run is based on the weighted average over three bins around the peak, approximately matching the 2.5-mm electrode pitch. The resulting slope of the linear fit establishes that the panel was able to reproduce the beam position. The PPS used for this experiment was similar to the panel shown in Fig. 2, having 128 columns (i.e., HV cathodes) and 32 rows (i.e., sense anodes), a Ni electrode pixel pitch of 2.5 mm, and a 0.40-mm gas gap that was filled with 1% $CO_2$ in 99% Ar at 600 Torr.

The steps observed in Fig. 10 (right) data are presumed to be caused by the intense beam saturating the central pixels. This saturation derives from the deliberately long time constants chosen for this first proton beam test. To further look into PPS saturation, we investigated the saturation response in a two-source experiment as follows: Four adjacent 32-cm-long horizontal signal readout lines (as shown in Fig. 11) were connected to discriminators whose outputs were OR'ed, and then, their combined signal rates were measured with a rate counter. High voltage was applied to two transverse Ni electrodes at varying distances from one another. Specifically, HV was applied always to one fixed line (#109 in Fig. 11), whereas the second line receiving high voltage was allowed to vary from #100 up to #109. The intersections of the isolated HV electrodes with four readout electrodes constituted the active pixels in this test. Each set of four pixels was exposed at first separately, then simultaneously to two partially collimated radioactive sources ($^{90}$Sr and $^{106}$Ru), yielding approximately similar rates of betas entering the gas gap region. These sources were positioned, one below the panel and one above, over the active pixels as indicated by the shaded regions in Fig. 11. The second source position was incremented from left to right across the panel. As in the proton beam test, a large quench resistance was deliberately selected to produce long cell recovery times and induce saturation along the high voltage line. The source-induced signal rates for the two respective groups of pixels were measured independently and then simultaneously. The result of this test is shown in Fig. 12. The rates of the two sources are observed to sum linearly over nearly the entire width of the panel. That is, the total readout rate of the two sources as applied simultaneously equals the sum of the rates of the sources applied separately. Deviations from this are observed only when the second source is brought very near the already saturated line #109. For example, when the second source is at line #107, the combined signal is still almost 90% of the sum of the two signals. However, we know from the beta simulation discussed earlier for Fig. 9 that our beta sources significantly scatter in the glass and thus disperse approximately one to two lines on either side of the source position. Thus, when the second source is brought up to line #107, the betas from the two sources can significantly overlap each other. Because the experiment started with line #109 already saturated, the additional signal overlap from the second source on line #107 cannot cause the combined beta signal on the sense lines in the vicinity of line #109 to proportionately increase, hence the disparity observed between the sum of the two signals versus the signal for both sources together (i.e., the pink and blue curves in Fig. 12). In summary, the initial experimental results of the double radiation source tests indicate that saturation is quite limited in extent. We expect that the new PPS structures now being fabricated should further reduce the contribution of nearby lines to the discharge, thus significantly reducing the amount of capacitive coupling and the degree of saturation.

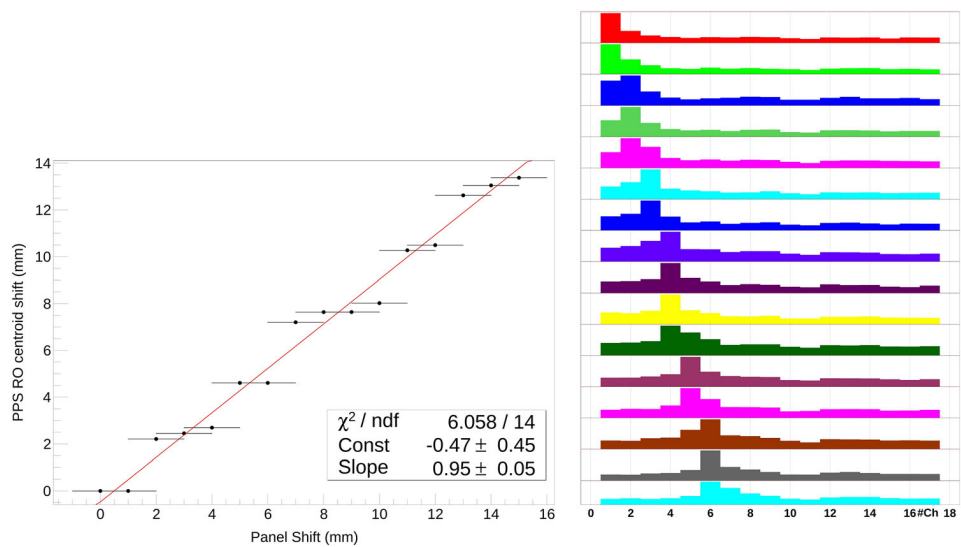

**FIGURE 10** — Position scan measurements with intense proton beam through 1-mm aperture.

Cosmic ray muons have allowed us to test the panel's response to MIPS. Using the setup described earlier (Fig. 3), we are able to associate signals induced in the panel with cosmic muons. In two different gas mixture experiments, we have measured the panel total efficiency over a 16-pixel area (i.e., $4 \times 4$ matrix) to be on average ~8%. We note that the active area pixel fill factor in this panel was only 11%. The total efficiency is defined as the ratio of signals in the panel that coincide with the trigger versus the total number of triggers (i.e., all of the triggers from the hodoscope are associated with cosmic ray muons). When taking into account that only the pixel area itself is active, it yields that per pixel, the efficiency to detect muons should be much higher at ~70%. In future panel designs, we expect to achieve active area pixel fill factors in the range of about 80%, thus representing a sevenfold increase in device efficiency. This efficiency can be further increased by an improved cell design combined with a larger gas gap. These anticipated enhancements should lead to high device efficiencies.

We have been developing a practical modeling and simulation capability for the purpose of providing (1) better insight into the device discharge characteristics and a clearer understanding of the interplay between the various device design parameters, materials selection, and electronics readout design; and (2) future design guidance for device optimization with respect to specific applications and to better understand the various performance tradeoffs associated with each particular device design. Our approach starts with a simplified schematic of a single PPS discharge cell. We then created a more realistic model and schematic of the discharge cell that includes stray capacitances, line resistance, and self-inductance. The parameters were determined from a COMSOL[8] electrostatic model. Finally, we expanded the single cell to a chain of cells by adding in the neighboring cells to form a larger array system. Represented in this expanded cell array/schematic are the embedded cell resistances, the cell capacitances, stray capacitances, self-inductances, and the termination resistance. The various capacitive couplings were modeled with COMSOL. COMSOL-3D[8] was employed to model the electric field and the charge motion inside the pixels, and the electronic properties of the different components (e.g., capacitances and inductances of the cells). SPICE was employed to simulate the electrical characteristics of the signal induced in the panel during discharge. The parameters in the SPICE models were determined with our COMSOL electrostatic model.

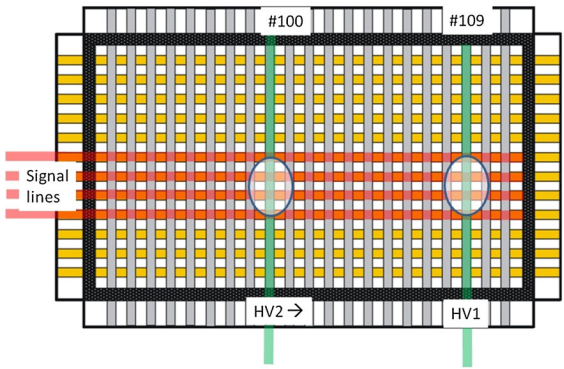

**FIGURE 11** — Double source test configuration. Shaded regions show approximate location of radioactive beta sources. The line labeled HV2 is incremented from left to right towards HV1.

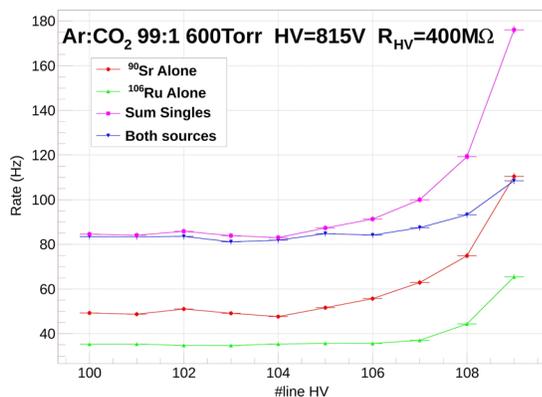

**FIGURE 12** — Results of two-source saturation test scan (see text).

The full SPICE model connects all of the neighboring cells into a single matrix to form a large cell array or a small panel sector. The full SPICE model allows us to test/evaluate the role of stray capacitance and inductances. We are now able to superimpose measured (i.e., experimental) signals over the SPICE simulations, resulting in an excellent match of the basic discharge shape (Fig. 13). By testing the influence of the various parameters, we are able to enhance our understanding of how these devices operate, their performance advantages and limitations, and how they can be optimized for specific applications. Perhaps the single most important insight gained by our modeling program is the critical role played by capacitive coupling to neighboring cells.

## 4 Technology projections and applications

We have used GEANT4 to model both particle and photon scattering for a variety of different PPS detector applications. The simulations have been encouraging and have led to interest in developing high-resolution, ultra-low mass PPS devices for various beam applications ranging from radiation therapeutics (e.g., hadron particle and/or gamma-ray therapy) to nuclear physics. We previously demonstrated that the PPS can detect ultraviolet photons at 366 nm,[3] as well as gamma rays from radioactive sources.[1] With the addition of an internal photocathode conversion layer, the PPS could potentially be the low cost, flat detector required for a variety of Compton telescope imaging applications.[9] By employing a gadolinium (Gd) conversion layer, the PPS could be configured as a neutron detector. Gadolinium with the highest neutron absorption coefficient of any element would function somewhat analogous to a photocathode, but instead of absorbing a photon and releasing a photoelectron, it would absorb a neutron and release a conversion electron into the gas to be detected by the PPS. Our GEANT4 simulations suggest that a low-mass Gd-foil-based PPS has the potential to be an efficient, nearly "gamma-blind" thermal neutron detector with a gamma-neutron discrimination ratio close to that required for $^3$He replacement neutron detectors.[2] An additional advantage of such a detector is that its low mass and thin profile could potentially yield an extremely light weight, compact package suitable for both large area installations as well as portable applications. Such devices could be particularly important for the detection of neutrons emitted by fissile materials (e.g., plutonium or enriched uranium) and thus for the homeland security market.

We have shown that the PPS is capable of detecting protons in intense beams with a position resolution consistent with the pixel granularity of our prototype device and in the energy range used for treating cancer. The detection of MIPS, such as cosmic-ray muons, has been demonstrated and is an important step in developing a PPS detector useful in high-energy physics applications. Similarly, the detection of radioactive ion beams using ultra-thin, low-mass PPS substrates is important to nuclear physics. Although we have focused primarily on ionizing particle detection, we are also interested in high-energy ionizing photon detection, such as for homeland security and X-ray radiation therapeutics. We have run simulations using GEANT4 on a PPS configured for high-energy X-ray detection (i.e., 6–8 MeV) and have found that a PPS-type device should be able to measure the incident beam in real time, as the patient is being treated with very little scattering of the beam. In the future, we plan to also explore applications for medical imaging such as positron emission tomography, computed tomography, and single-photon emission computed tomography. Given the large scope of possible applications and the cost advantages of the PPS technology, the commercial impact and potential benefits of this technology could have a large impact on a number of important fields. Radiation detectors for homeland security and various medical applications constitute a multibillion dollar business opportunity. Unlike high-resolution flat panel displays that are consumer electronics, "the PPS when fully developed should be able to sell for one to two orders of magnitude above its manufacturing price and still be priced well below competing radiation detector technologies." For example, although PDPs currently sell for $0.02 per square centimeter, photomultiplier tubes sell for about $3 per square centimeter, solid state detectors at the low end sell in the range of $30 per square centimeter but can vary widely depending upon the semiconductor material, and multichannel plate detectors sell for more than $300 per square centimeter.

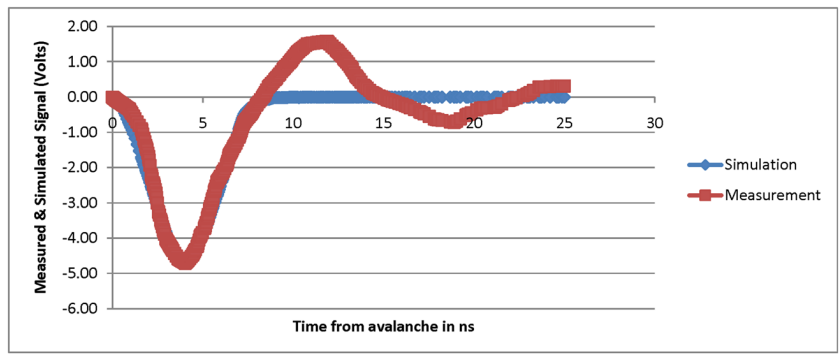

**FIGURE 13** — Modeling with SPICE and COMSOL. The red line is an experimental sense electrode signal, and the blue line is our simulated signal for the same panel.

Having tested PPS panels with several beta sources as well as cosmic-ray muons, we observe "hits" from such ionizing particles when they enter the voxel space defined by the cell discharge gap volume dimensions. To take advantage of this, new cell structures have been designed to significantly increase the effective cell active discharge region to maximize device efficiency. We are also modifying our device fabrication process and the cell configuration to improve pixel uniformity and thereby increase the operational range for the panel. These enhancements are intended to lead to PPS devices with order of magnitude higher spatial resolution and reduced jitter. We feel that such projections are realistic because we have already demonstrated the potential for the PPS to achieve both high position and temporal resolution.

## 5 Conclusions

The PPS was conceived to benefit from the mature PDP technology base with its low-cost manufacturing infrastructure by using similar materials and manufacturing processes. Thus, in addition to offering inexpensive materials and fabrication processes for the production of highly pixelated, high-performance devices, the PPS offers a number of other potential advantages including pulse rise times of 1–2 ns, pulse widths (full width at half-maximum) on the order of 2 ns with a temporal response or timing jitter of ~5 ns, high position resolution, low power consumption (e.g., ~20 µW/cm$^2$ at a "hit" rate of 20 kHz/cm$^2$), high internal gain (e.g., ~10$^7$ gain for 1-mm pixel), a thin and compact flat panel structure with low mass, a hermetic seal eliminating the need for a gas flow system, and a materials composition that is inherently radiation-damage resistant (e.g., glass substrates, metal electrodes, and stable gas mixtures). These potential PPS attributes are attracting significant interest for applications ranging from detection of nuclear materials for homeland security, proton beam detectors for improved radiation therapeutics in treating cancer as well as proton imaging in real time and proton dosimetry, detecting MIPS for high-energy physics, radioactive ion beams monitors for nuclear physics, medical imaging, and others.

We have been able to demonstrate that for a given panel structure and gas, the discharge signals look remarkably uniform and are inherently digital. We expect faster discharge times in the sub-nanosecond range with high position resolution as we transition to smaller cell sizes, better cell physical and electrical isolation including minimal if any discharge spreading, and lower panel capacitance. We believe that the fast rise times and short pulse durations are largely due to the very high gain of the PPS Geiger-mode electron avalanche.

Key objectives of our initial PPS experimental program were to demonstrate that (1) such devices can be fabricated as high-gain, micropattern detectors and successfully operated beyond the proportional region and above the gas breakdown voltage as Geiger-mode-type devices with high-performance capability; (2) PPS discharges can be made to self-terminate and can be self-contained to a localized cell site to yield both high spatial and high temporal resolution; (3) low-cost, commercial PDP technology can be modified to detect ionizing radiation; (4) signals have fast discharge times and large amplitudes; (5) hermetically sealed PPS gas devices appear to be stable; (6) useful models can be constructed with simulations that can be experimentally tested to confirm and enhance our understanding of how these devices operate and to evaluate optimization strategies. In the future, we hope to use these models to investigate the performance advantages and limitations of new PPS designs for specific applications. We are gratified that all six of the initial program objectives have been confirmed, and we are now moving to focus on specific device applications and commercialization. In summary, we believe that we have been able to demonstrate the viability, merit, and potential capability of the PPS as a hermetically sealed, high-gain, rad-hard detector with both high spatial and high temporal resolution, high rate capability, and low cost.

### *Acknowledgments*

This work was supported in part by the US Department of Energy under grant numbers DE-FG02-07ER84749, DE-SC0006204, DE-SC0006219, and DE-FG02-12ER41788. This work was also partially supported by the Office of Nuclear Physics at the US Department of Energy and the United States–Israel Binational Science Foundation under grant number 2008123.

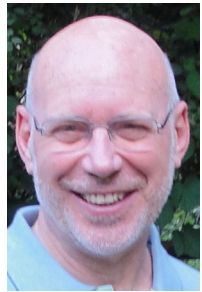

**Peter Friedman** received his PhD degree from the University of Michigan in physical chemistry in 1974. He continued at Michigan as a post-doctoral scholar for 2 years in the field of solid state spectroscopy and energy transfer. He moved on to eventually become director of research and president of Photonics Imaging, Inc. in 1989, which was the first company in the world to demonstrate a full-color AC-PDP for NTSC definition TV (525 lines/30 frames per second) in 1991 and a full-color 21-in. diagonal $1024 \times 1280$ workstation monitor in 1995, which is still today the finest pixel pitch PDP product ever made. From 1996 to 2004, he served as chairman of the board and chief technology officer of Photonics Systems, Inc. of Northwood, Ohio. In 2004, he founded Integrated Sensors, LLC of Ottawa Hills, Ohio, and has since served as president and chief technology officer. His most recent accomplishments include the invention of the plasma panel radiation detector for which he has received 11 patents.